\documentclass[10pt]{article}
\usepackage[left=10mm,right=10mm,top=15mm,]{geometry}
\usepackage{amsfonts,amsmath,amsthm,amssymb}
\usepackage{multicol} % allows figures to span all columns
\usepackage[inline]{enumitem}
\usepackage{graphicx}
\usepackage{caption}    % array of figures
\usepackage{subcaption}
\usepackage{url}
\usepackage{float}

\def\figs{Figures}   % path to image files

\title{\sf Linux-Tomcat Application Performance on Amazon AWS \vspace{-1cm}}
\date{\small Manuscript of \today}

\begin{document}
\maketitle

\begin{quote}
{\noindent \sf The need for Linux system administrators to do performance management
 has returned with a vengeance. Why? The cloud. Resource consumption in the cloud is all about pay-as-you-go. 
This article shows you how performance models can find the most cost-effective deployment of 
an application on Amazon's cloud. ---Neil Gunther and Mohit Chawla}
\end{quote}

\vspace{0.5cm}
 
\begin{multicols}{3}

%\section*{} No opening section head
\noindent 
While the advantages of off-premise cloud services are manifest---lower infrastructure capex, lift-and-shift migration,  versatile elastic capacity---the cloud also reintroduces fee-for-service: a forgotten concept in the Linux world that is all too familiar in the mainframe world. Once your application is deployed, pretty soon you are talking real money.

The magnitude of those cloud utility costs can come as a surprise, just like an unexpectedly big electricity bill. And that reinvigorates the need for sysadmins to undertake performance and capacity management, but this time, directed at cloud applications. 
% reinvigorate = give new life to but rejuvenate = to make young again  

In this article, we show you how, 
by combining production JMX (Java Management Extensions)~\cite{jmx} data with a custom performance model, it becomes feasible to determine the optimal configuration and Amazon Web Services (AWS)~\cite{aws} policies for 
an application running on a Linux-hosted Tomcat cluster~\cite{tomcat}. 
Once the performance model is established, it facilitates ongoing cost-benefit analysis of various EC2 Auto Scaling policies~\cite{cloudx}.

The entire application runs in the Amazon cloud. 
Smartphone users make requests to the Apache HTTP-server (versions 2.2 and 2.4) via 
Amazon's Elastic Load Balancer (ELB) on its Elastic Compute Cloud (EC2). 
The Tomcat thread-server (versions 7 and 8) running on an EC2 instance makes procedure calls to a variety of external services belonging to third parties, e.g, hotel booking and rental car services.
AWS Auto Scaling (A/S) controls the number of operational EC2 instances in the cluster, based on the amount of incoming smartphone traffic and the configured A/S policies. 
ELB balances incoming traffic across all the entire cluster.

\begin{figure*}[ht]
\centering
\begin{subfigure}{0.33\textwidth}
  \centering
  \includegraphics[scale=0.325]{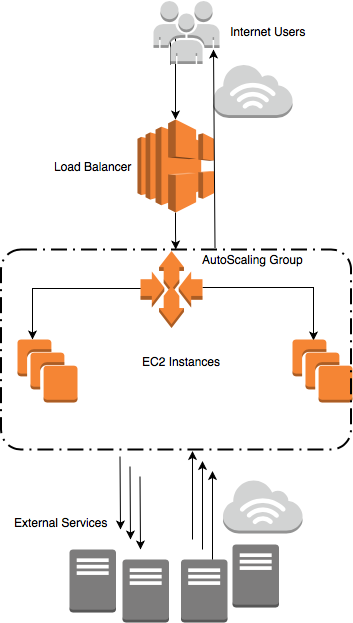}
  \caption{Auto Scaling grouping} \label{fig:apparch}
\end{subfigure}%
\begin{subfigure}{0.33\textwidth}
  \centering
  \includegraphics[scale=0.325]{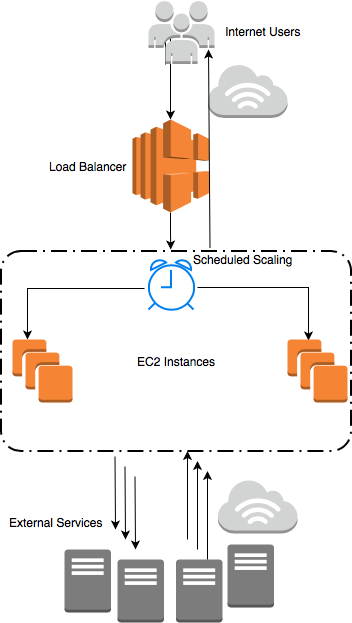}
  \caption{Scheduled Scaling grouping} \label{fig:awssched}
\end{subfigure}%
\begin{subfigure}{0.33\textwidth}
  \centering
  \includegraphics[scale=0.325]{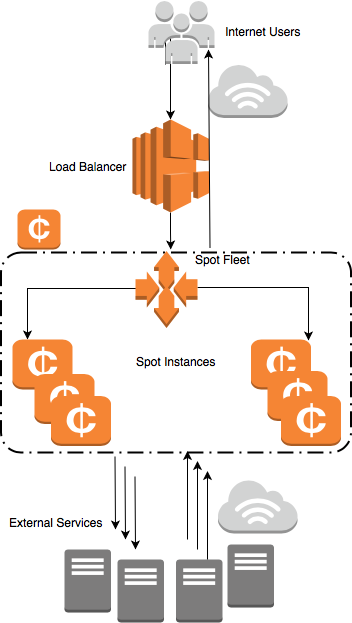}
  \caption{Spot Instance grouping} \label{fig:awsspot}
\end{subfigure}
\caption{AWS configurations with different performance attributes} \label{fig:awsapp}
\end{figure*}

\section*{\sf Application Architecture}
We only need to focus on the performance of a single EC2 instance in the AWS cluster.
The ELB is assumed to guarantee the other instances perform in the same way.

A high-level schematic is shown in Figure~\ref{fig:apparch}. 
Starting at the top, the major architectural components are:
\begin{enumerate}
\item Smartphone users on the Internet making requests to the application residing on AWS.
\item Incoming requests are load balanced across the AWS virtualized cluster.
\item The A/S group handles dynamic cluster sizing.
\item Tomcat servers running on EC2 instances in the A/S group.
\item Communication with third-party services that are external to the Tomcat application.
\end{enumerate}

\noindent
On any individual EC2 instance, an incoming HTTP request from a smartphone user is processed by the Apache HTTP server and Tomcat.
The Tomcat server then sends multiple requests to the external services which are selected based on the original smartphone request.
The external services respond and Tomcat computes some business logic based on each of their responses.
Tomcat then returns the final processed response back to originating  user's smartphone.

\section*{\sf Performance Tools}
Since Apache Tomcat is a Java application, we can make extensive use of the JMX interface to 
data from the JVM via:
\begin{itemize}
\item {\tt jmxterm}
\item VisualVM
\item Java Mission Control
\item Datadog {\tt dd-agent}
\item Custom scripts
\end{itemize}

\noindent
Other open-source data collection and performance analysis tools were also used (see Table~\ref{tab:tools}).

\begin{table*}[ht]
\begin{center}
\caption{Open source performance tools} \label{tab:tools}
\begin{tabular}{l|l}
\multicolumn{1}{c}{\bf OSS Tool} & \multicolumn{1}{c}{\bf Purpose}\\
\hline
Datadog  & Data monitoring that also integrates AWS CloudWatch metrics\\
Collectd & Collection of Linux performance statistics \\
Graphite & Collection and storage of application metrics \\
Grafana  & Interactive plotting of performance data as time series\\
R        & Statistical data analysis packages\\
RStudio  & IDE for R program development\\
PDQ      & {\em Pretty Damn Quick} performance modeling library\\
\hline
\end{tabular}
\end{center}
\end{table*}%

%\section*{\sf Production Data Collection}
Raw performance data was primarily collected by a Datadog {\tt dd-agent} and 
used to derive input metrics for  performance models written in PDQ.
Smartphone-user requests are treated as a homogeneous workload. 

JMX has a managed bean or {\em MBean} object called {\tt GlobalRequestProcessor} that provides two raw metrics: 
\begin{enumerate*}[label=(\roman*)]
\item {\tt requestCount}: total number of requests ($C$),   
\item {\tt processingTime}: total processing time for all requests ($T_{proc}$).
\end{enumerate*}

The average requests per second, $X_{dat}$, is derived by converting {\tt requestCount} to a rate in the {\tt datadog} configuration. 
The average response time, $R_{dat}$, is derived from
\begin{equation*}
R_{dat} = \bigg( \dfrac{T_{proc}}{T_{sample}} \bigg) \bigg( \dfrac{T_{sample}}{C} \bigg)
\end{equation*}
in data in each sample interval, $T_{sample}$.

We now apply two variants~\cite{njgblog} of Little's law
\begin{enumerate*}[label=(\alph*)]
\item the {\em macro}-scopic version $N = X * R$, which is defined in terms of the larger-scale response time $R$, and  
\item the {\em micro}-scopic version $U = X * S$, which is defined in terms of the smaller-scale service time $S$.
\end{enumerate*}
These definitions~\cite{linmag, pdqbook}  are used to derive additional performance metrics from the raw data, as follows.

\begin{enumerate}
\item  The macroscopic law produces:
    \begin{itemize}
    \item The estimated number of concurrent requests, $N_{est}$, in Tomcat during each measurement interval.
    \item Verification that $N_{est}$ is identical to the measured number of threads $N_{dat}$ in the Tomcat server.
    \end{itemize}
    
\item  The microscopic law produces:
  \begin{itemize}
  \item The measured processor utilization, $U_{dat}$, reported by the {\tt dd-agent}, as a decimal fraction, not a percentage. 
  \item Confirmation that the throughput collected in JMX data is the same as $X_{dat}$.
  \item The estimated service time metric, $S_{est} = U_{dat} / X_{dat}$.
  \end{itemize}
\end{enumerate}

\noindent
The result of all this data extraction is to yield the small number of performance metrics in Table~\ref{tab:metrix} needed to parameterize our performance models.

The Unix epoch {\tt Timestamp} interval between rows in Table~\ref{tab:metrix} is 300 seconds~\cite{cloudx}. 
Once again, Little's law confirms the relationships between these metrics:
\begin{enumerate*}[label=(\roman*)]
\item $N_{est} = X_{dat} * R_{dat}$ from macroscopic Little's Law  and 
\item $U_{dat} = X_{dat} * S_{est}$ from microscopic Little's Law,
\end{enumerate*}
each time-averaged over the sampling interval $T_{sample} = 300$ seconds.

\begin{table*}
\centering
{\small 
\begin{verbatim}
      Timestamp          Xdat            Nest            Sest          Rdat          Udat
      1486771200000      502.171674      170.266663      0.000912      0.336740      0.458120
      1486771500000      494.403035      175.375000      0.001043      0.355975      0.515420
      1486771800000      509.541751      188.866669      0.000885      0.360924      0.450980
      1486772100000      507.089094      188.437500      0.000910      0.367479      0.461700
      1486772400000      532.803039      191.466660      0.000880      0.362905      0.468860
      1486772700000      528.587722      201.187500      0.000914      0.366283      0.483160
      1486773000000      533.439054      202.600006      0.000892      0.378207      0.476080
      1486773300000      531.708059      208.187500      0.000909      0.392556      0.483160
      1486773600000      532.693783      203.266663      0.000894      0.379749      0.476020
      1486773900000      519.748550      200.937500      0.000895      0.381078      0.465260
      ...
\end{verbatim}
}
\caption{Distilled performance metrics}  \label{tab:metrix} 
\end{table*}

\begin{figure*}[ht]
\centering
\begin{subfigure}{0.5\textwidth}
  \centering
  \includegraphics[scale=0.5]{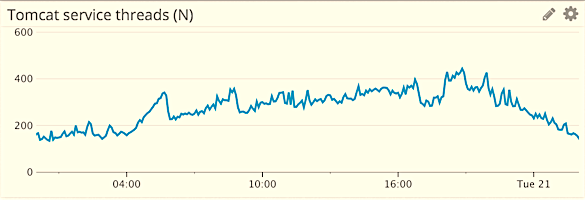}
  \caption{Time-dependent monitored concurrency} \label{fig:jmxN}
\end{subfigure}%
\begin{subfigure}{0.45\textwidth}
  \centering
  \includegraphics[scale=0.425]{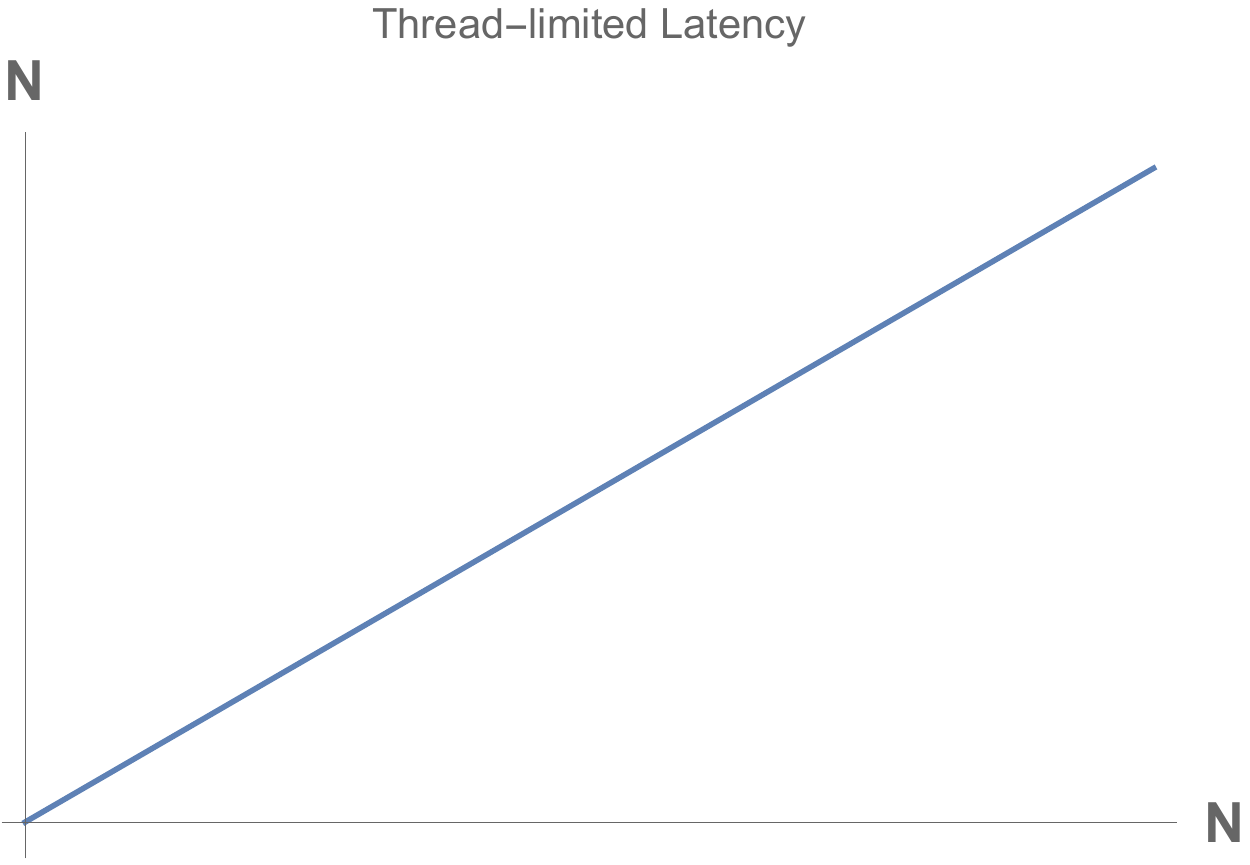}
  \caption{Time-independent concurrency} \label{fig:pdqN}
\end{subfigure}
\begin{subfigure}{0.5\textwidth}
  \centering
  \includegraphics[scale=0.5]{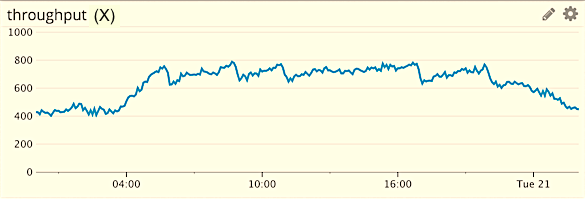}
  \caption{Time-dependent monitored throughput} \label{fig:jmxX}
\end{subfigure}%
\begin{subfigure}{0.45\textwidth}
  \centering
  \includegraphics[scale=0.425]{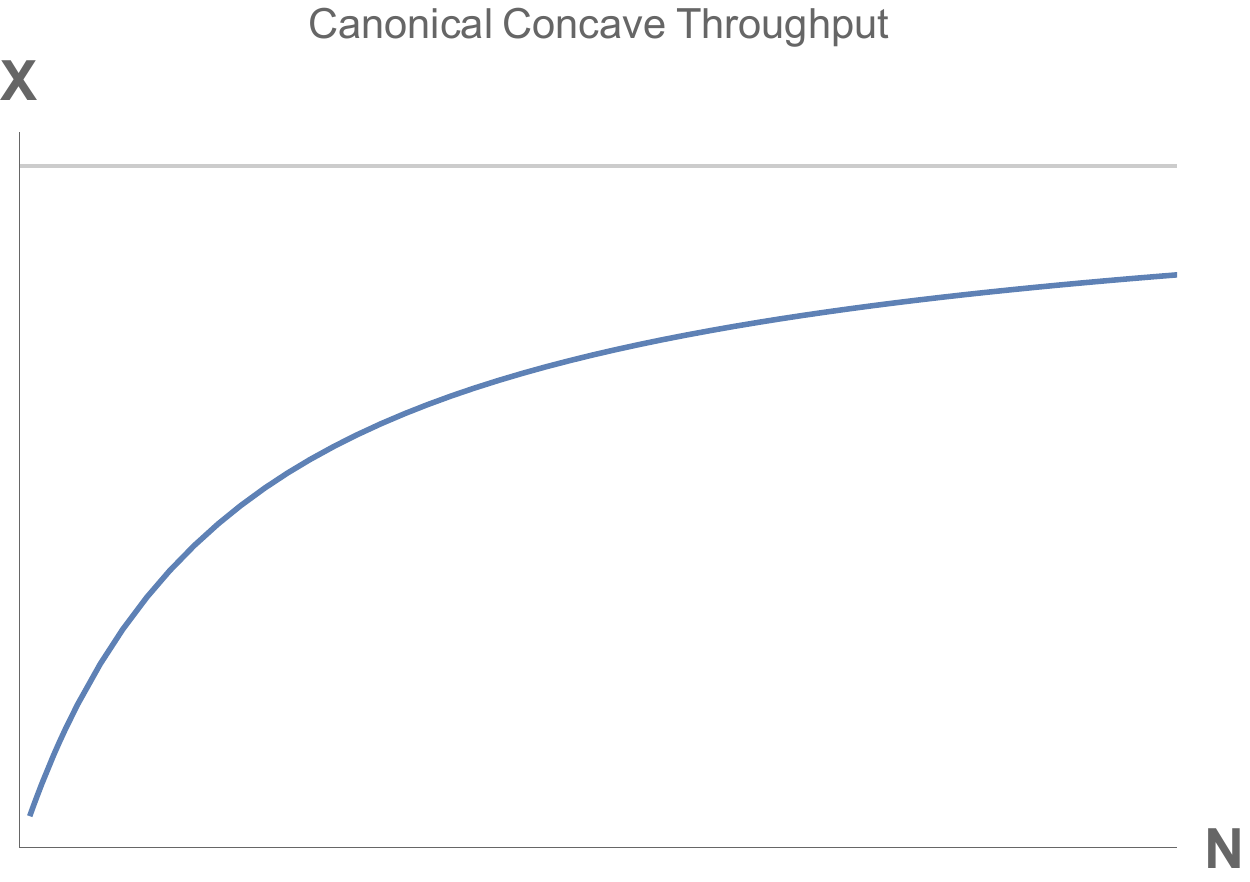}
  \caption{Time-independent throughput} \label{fig:pdqX}
\end{subfigure}
\begin{subfigure}{0.5\textwidth}
  \centering
  \includegraphics[scale=0.5]{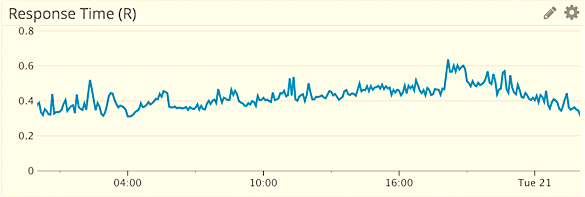}
  \caption{Time-dependent monitored response time} \label{fig:jmxR}
\end{subfigure}%
\begin{subfigure}{0.45\textwidth}
  \centering
  \includegraphics[scale=0.425]{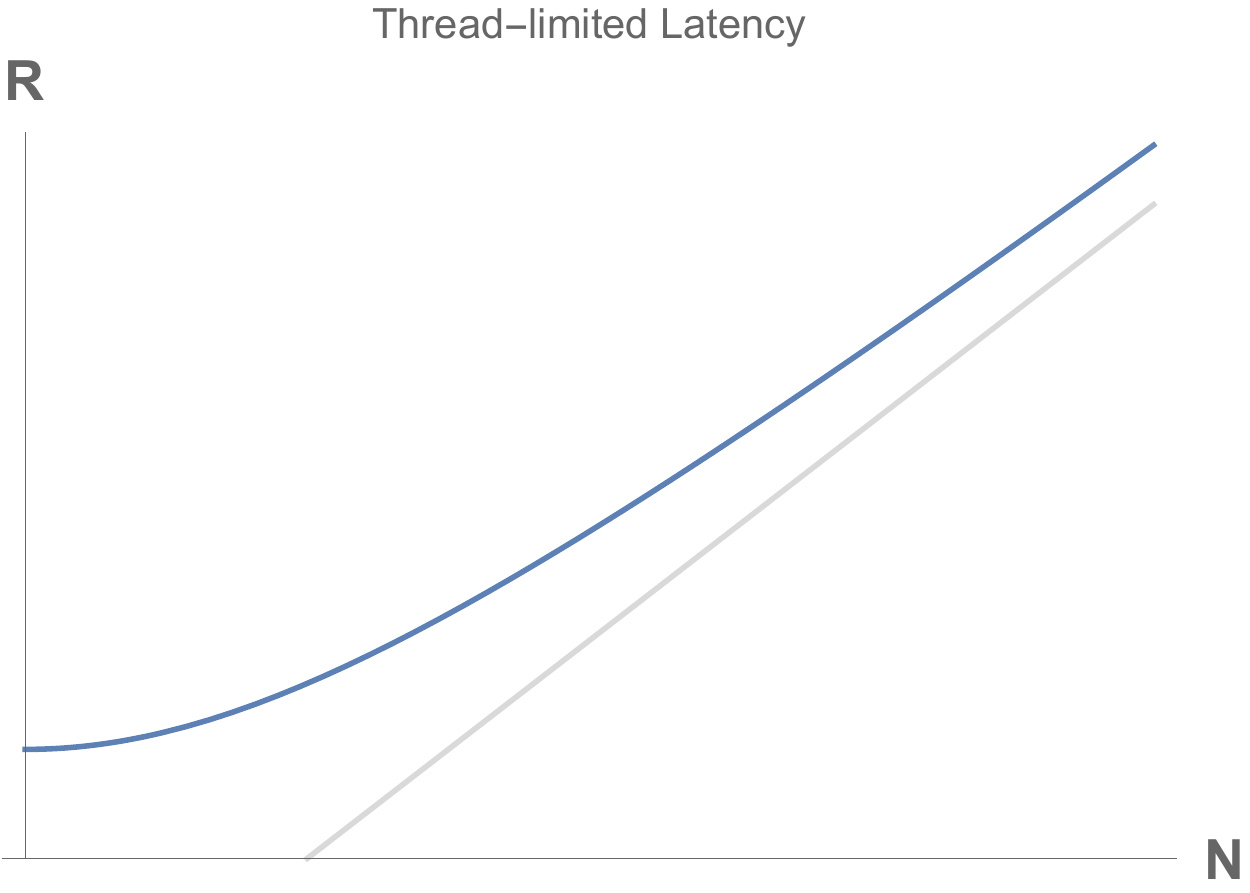}
  \caption{Time-independent response time} \label{fig:pdqR}
\end{subfigure}
\caption{Comparison of monitored metrics (left) with their steady-state counterparts (right)}
\label{fig:metrix}
\end{figure*}

\section*{\sf Different Data Views}
Figures~\ref{fig:jmxN},~\ref{fig:jmxX}, and~\ref{fig:jmxR} show of performance metrics as they are seen in a typical monitoring tool, i.e., a time-{\em dependent} view with 
concurrency ($N$), throughput ($X$) and response time ($R$) on the y-axis and time on the x-axis. 
This view shows how each metric evolves and therefore when certain values occurred. 

In contrast to this canonical view of monitored data,  
Figures~\ref{fig:pdqN},~\ref{fig:pdqX}, and~\ref{fig:pdqR} 
show the time-{\em independent} or steady-state view of concurrency, throughput and response time as a function of load 
on the x-axis. 

This steady-state view is achieved by a transformation that selects a particular point in time in the monitored data and takes the corresponding value of $X$ and $R$ at that same time. In this way, time becomes implicit.

The steady-state throughput profile, $X$, is a {\em concave} function in Figure~\ref{fig:pdqX}. Think of it as forming a 
``cave'' with respect to the x-axis. 
It starts out rising almost linearly at low loads, but eventually forms a plateau when the utilization of a key resource saturates at 100\% busy and forms a bottleneck.

Conversely, the steady-state response time profile, $R$, is a {\em convex} function in Figure~\ref{fig:pdqR}.
It starts out as a plateau at low loads, but rises linearly in the saturation region due to increasing queue lengths. For this reason, the response-time shape is often referred to as the ``hockey stick'' curve~\cite{pdqbook}.

In a nutshell, once we have the steady-state view of production data, queueing theory tells us what to expect.

%%%%%%%%%%%% PDQ in R sidebar  %%%%%%%%%%%% 
\begin{figure*}[ht]
\centering
\fbox{\begin{minipage}{0.45\textwidth}
{\bf \sf \large PDQ-Bibliothek in R}\\
\footnotesize
\sf PDQ (Pretty Damn Quick) ist ein Modellierungstool
für die Analyse der Performance von
Rechner-Ressourcen wie Prozessoren, Platten
oder Gruppen von Prozessen, die diese Ressourcen
beanspruchen. Ein PDQ-Modell wird
mit Hilfe von Algorithmen aus der Queueing-
Theorie analysiert. Die aktuelle Release erlaubt
das Erstellen solcher Performance-Modelle
in C, Perl, Python und R.
Die Beispiele dieses Artikels benutzen Funktionen
wie folgende:
\begin{itemize}
\item {\tt pdq::Init()} initialisiert interne PDQ-Variable.
\item {\tt pdq::CreateOpen()} erzeugt einen Workload.
\item {\tt pdq::CreateNode()} erzeugt einen Server.
\end{itemize}
\end{minipage}
\hspace{0.025\textwidth}
\centering
\footnotesize
\begin{minipage}{0.45\textwidth} \sf
\begin{itemize} 
\item {\tt pdq::SetDemand()} setzt die Workload-Servicezeit der Server-Ressource.
\item {\tt pdq::Solve()} berechnet Performance-Metriken.
\item {\tt pdq::Report()} erzeugt einen generischen Report.
\end{itemize}
Weitere Informationen finden sich auf der\\
Website \url{www.perfdynamics.com/Tools}\\
\"{U}berblick: \url{...com/PDQ.html}\\
Download: \url{...com/PDQcode.html}\\
Manual: \url{...com/PDQman. html}\\

Der Autor dieses Artikels und Paul J. Puglia entwickeln und pflegen PDQ seit 2012.
\end{minipage}
}
\end{figure*}
%%%%%%%%%%%% %%%%%%%%%%%% %%%%%%%%%%%% 

\section*{\sf Performance Models} 
In the subsequent sections, we are going to analyze this 
cloud application by representing it with the PDQ performance modeling tool~\cite{pdqtool} 
in the R language. (See sidebar {\bf \sf PDQ-Bibliothek in R}) 

More detailed background, with examples that show you how to use PDQ in Perl and Python, can be found in 
previous editions of Linux-Magazin~\cite{linmag} and Linux Technical Review~\cite{berechenbare}.

Because of R's ability to ingest diverse data formats (either as text or database queries), apply a huge number of robust statistical routines to those data, extract PDQ modeling parameters from those data, execute the corresponding PDQ models, and visualize numerical results graphically---all from within the same script---R is our goto weapon for performance analysis.

One might reasonably wonder if PDQ can be applied at all to modern cloud architectures. 
In queue-theoretic terms, a modern computer system can be thought of as a directed
graph of individual buffers where requests can wait for service at a shared 
computational resource, e.g., a locally attached disk. Since a
buffer is just a queue, all computer systems can be represented as a directed
graph of queues. The directed arcs represent flows between different queueing resources.
PDQ computes the performance metrics of such a graph. 
A directed graph of queues is sometimes referred to as a {\em queueing network model}~\cite{pdqbook}.

\begin{figure*}[ht]
\centering
\begin{subfigure}{0.5\textwidth}
  \centering
  \includegraphics[scale=0.45]{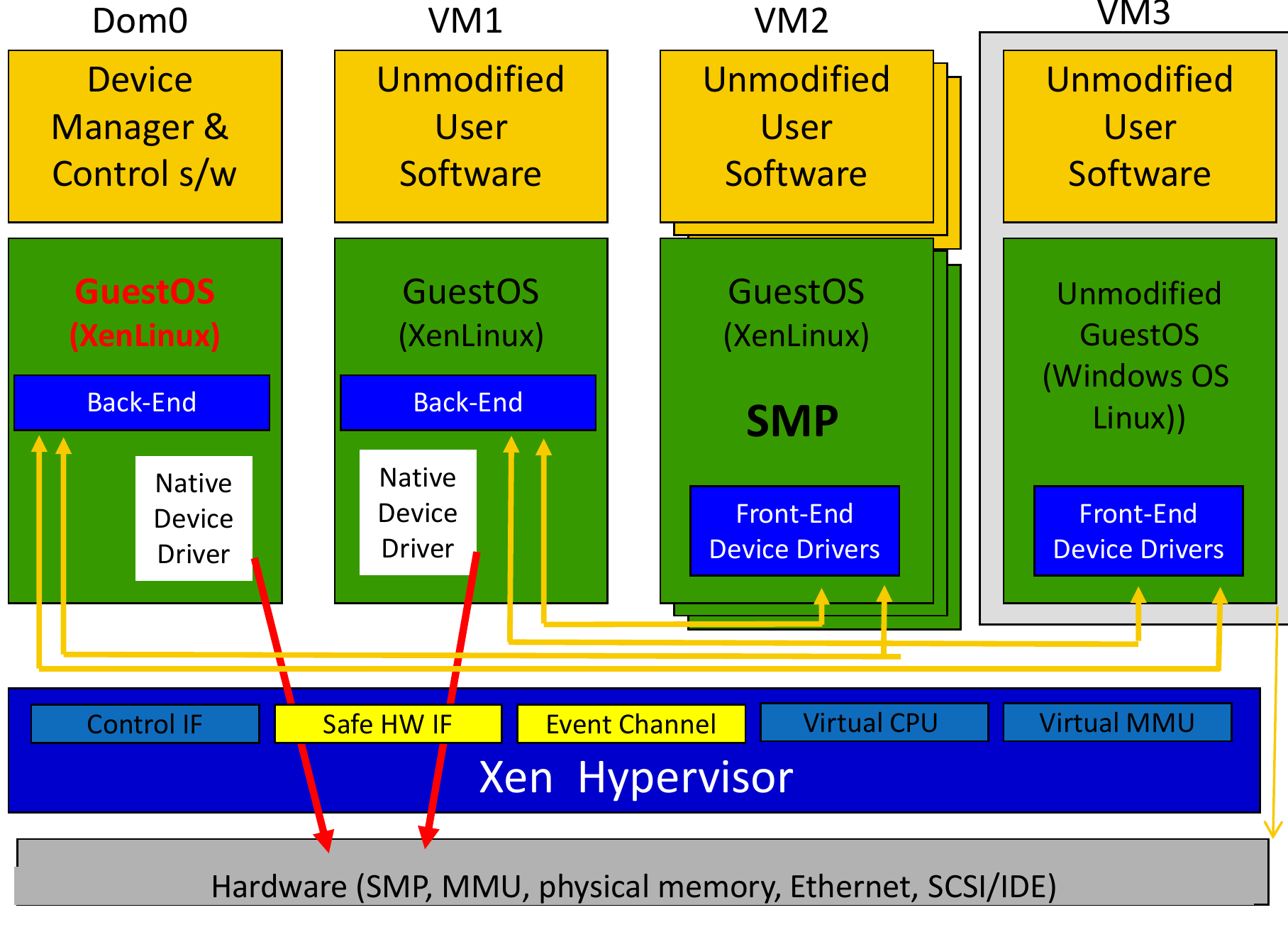}
  \caption{Xen hypervisor architecture (Source~\cite{xenintel})} \label{fig:xenarch}
\end{subfigure}%
\hspace{0.075\textwidth}
\begin{subfigure}{0.35\textwidth}
  \centering
  \includegraphics[scale=0.72]{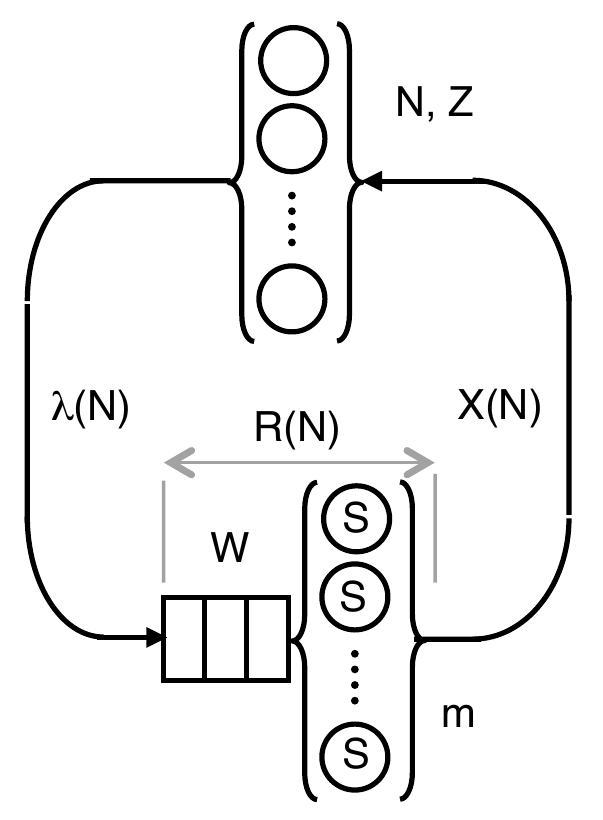}
  \caption{Queueing model of Tomcat threads} \label{fig:pdqtc}
\end{subfigure}%
\caption{XenLinux VM instances (left) and queueing model of Tomcat instance (right)} \label{fig:fstc}
\end{figure*}

From the performance modeling standpoint, the cloud is just a collection of virtualized servers on the other end of the Internet~\cite{xencamb,xenaws,vmware}.
Moreover, most virtualization involves a {\em hypervisor}, such as Xen Server in Figure~\ref{fig:xenarch}.
Hypervisors employ a {\em fair-share} scheduler~\cite[Chap. 7]{gcapbook}. 
In case there is any doubt, the documented default share allocation per VMware ``guest'' (i.e., a virtual machine instance) is 1,000 shares~\cite{vmware}. 

A fair-share scheduler is necessary in order to juggle each of the virtual machine instances,
each of which contains their own instance of Linux or some other operating system scheduler, 
across the underlying computational resources (especially bare-metal processors). 
AWS {\em CloudWatch} provides performance metrics as seen by the hypervisor~\cite{cloudwatch,cloudwatch2}.

On the spectrum of operating systems, a fair-share scheduler falls somewhere between a {\em batch} scheduler (found in data-processing mainframe computers) and a standard {\em time-share} scheduler (such as Linux).

In order that is can meet its execution window, a batch scheduler does not allow for any processing interruptions, whereas a time-share scheduler is all about interruptions (via a time quantum) in order that each user thinks he or she is the only user active on the system~\cite{gcapbook}.

Even with all these different schedulers at play, the performance model of the Tomcat application turns out to be far simpler than one might expect. 
Although the A/S group in Figure~\ref{fig:apparch} consists of multiple EC2 instances,
they are all just a replication of the same Tomcat application. Moreover, they are all doing the same thing. If that were not true, it would mean the ELB load balancer was not doing its job correctly and that would 
constitute a functional failure. 

From this, it follows that we only need to model a single EC2 instance to know how the entire AWS cluster should be performing. 

The PDQ queueing representation of the Tomcat application on a single EC2 instance is shown in Figure~\ref{fig:pdqtc}.
Starting at the top of the diagram, we identify the following components symbolically:
\begin{enumerate}
\item 
$N$ user requests are denoted schematically as the round ``bubbles'' within the curly braces. 
Over a twenty four hour period, $N$ typically can range between 100 and 500 requests. 
$N$ never reaches zero because there is always some activity due to multiple time zones.
There can only be a {\em finite} number of requests in the system in any sample interval.
\item 
$Z$ denotes the user think time: the time between response to a previous request and the issuance of the next. 
A typical example is the delay between a user seeing a requested web page and eventually clicking a hyperlink on that rendered web page. 

In the case of our Tomcat application, the smartphone-users are external to Figure~\ref{fig:pdqtc}, so there is no formal think-time in our PDQ model of Tomcat threads, i.e., we set \mbox{$Z=0$}.
\item 
Moving down to the bubbles at the bottom of Figure~\ref{fig:pdqtc}, 
$m$ denotes the maximum number of Tomcat threads that are available to service user requests. 

In addition, and unlike the top bubbles, there can be requests waiting for service and this waiting line, or buffer, is indicated by the ``blocks'' to the left of the thread bubbles in Figure~\ref{fig:pdqtc}. See the {\sf Waiting Where?} section for more on this.
\item 

The average time it takes to service a Tomcat thread is denoted by $S$.
\item 
That service time, $S$, together with any incurred waiting time, $W$, corresponds to the average response time, \mbox{$R=W+S$}.
\item
The rate of requests arriving in the bottom thread queue is denoted by $\lambda$.
\item 
System throughput is denoted by $X$. We can reasonably assume that the EC2 instance is running in steady state 
(i.e., no significant transients) so that, $\lambda=X$
\end{enumerate}

\noindent
For our purposes, we will generally focus on the performance characteristics of $X$ and $R$ in response to the load $N$.
This functional dependence on the load is sometimes made explicit by writing $X(N)$ and $R(N)$.

In general, each user request gets a Tomcat thread as soon as it arrives into the system. 
Since there is no waiting time, $W$, in that case, the response time $R$ is the same as the service time $S$.
However, there can be more requests, $N$, in the system than available threads, $m$.

\section*{\sf PDQ Model Calibration} 
\begin{figure*}[ht]
\centering
\begin{subfigure}{0.33\textwidth}
  \centering
  \includegraphics[scale=0.325]{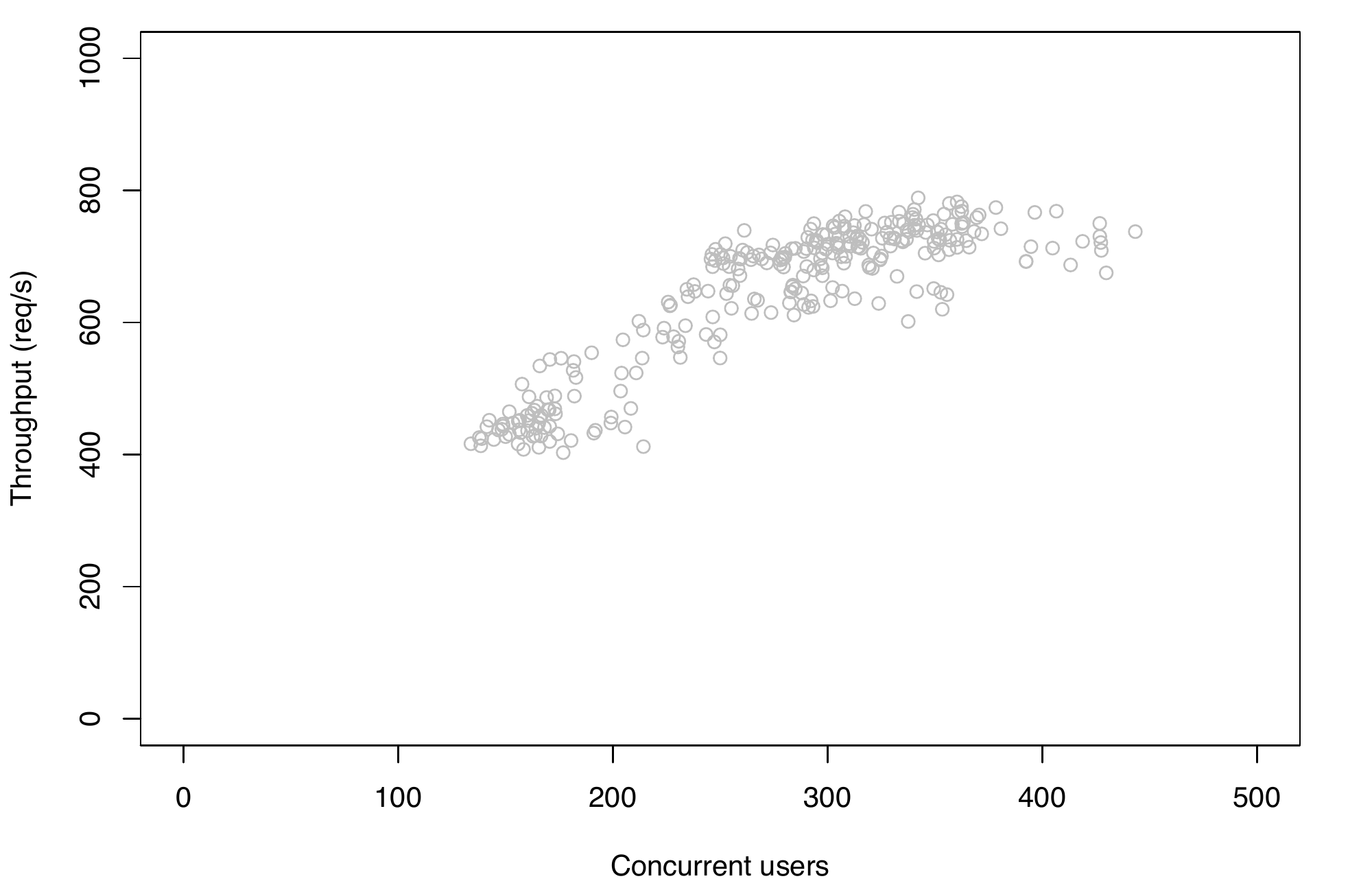}
  \caption{Production throughput data} \label{fig:jmxX2016jul}
\end{subfigure}%
\begin{subfigure}{0.33\textwidth}
  \centering
  \includegraphics[scale=0.325]{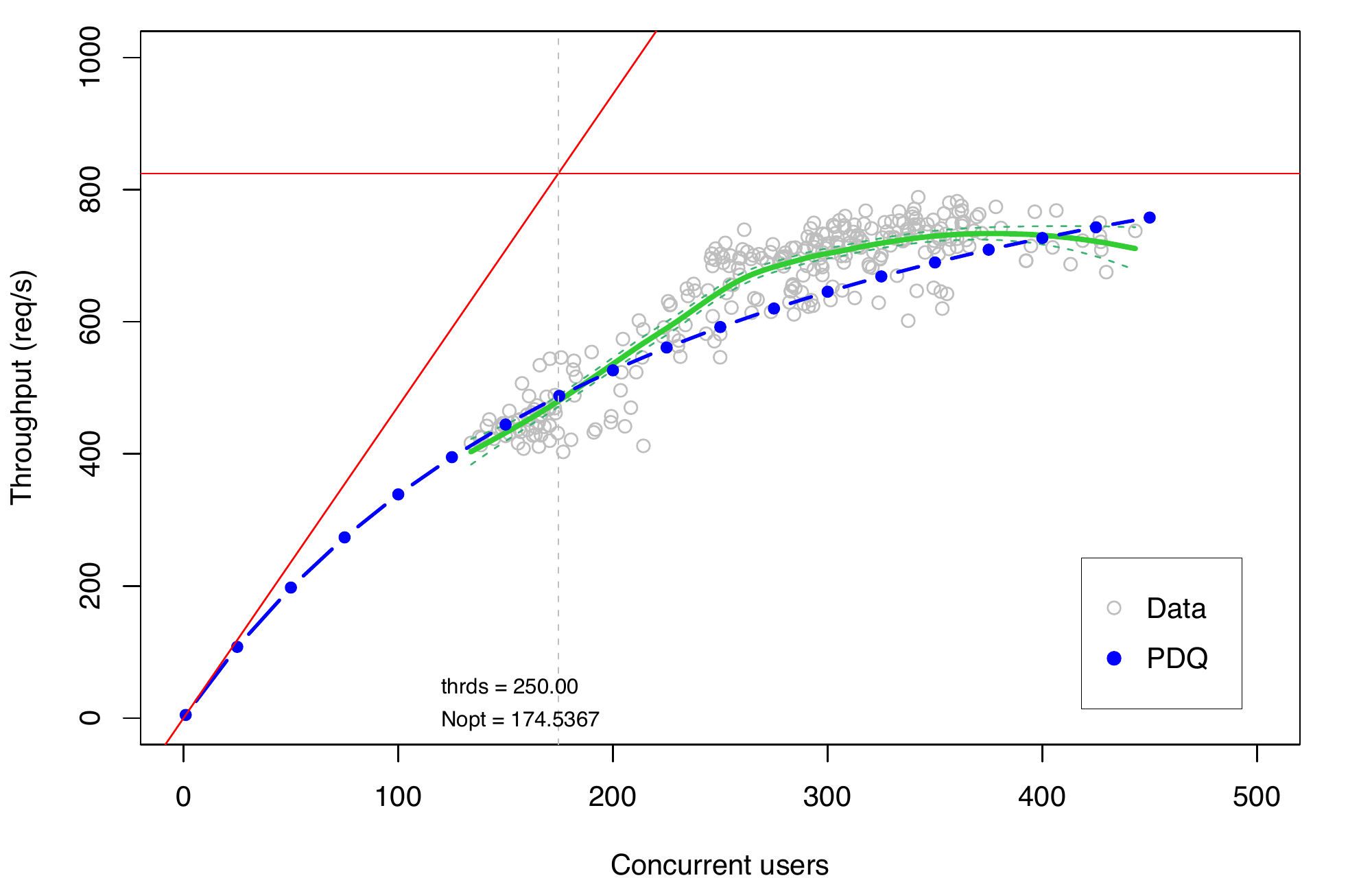}
  \caption{PDQ throughput model} \label{fig:pdqX2016jul}
\end{subfigure}%
\begin{subfigure}{0.33\textwidth}
  \centering
  \includegraphics[scale=0.325]{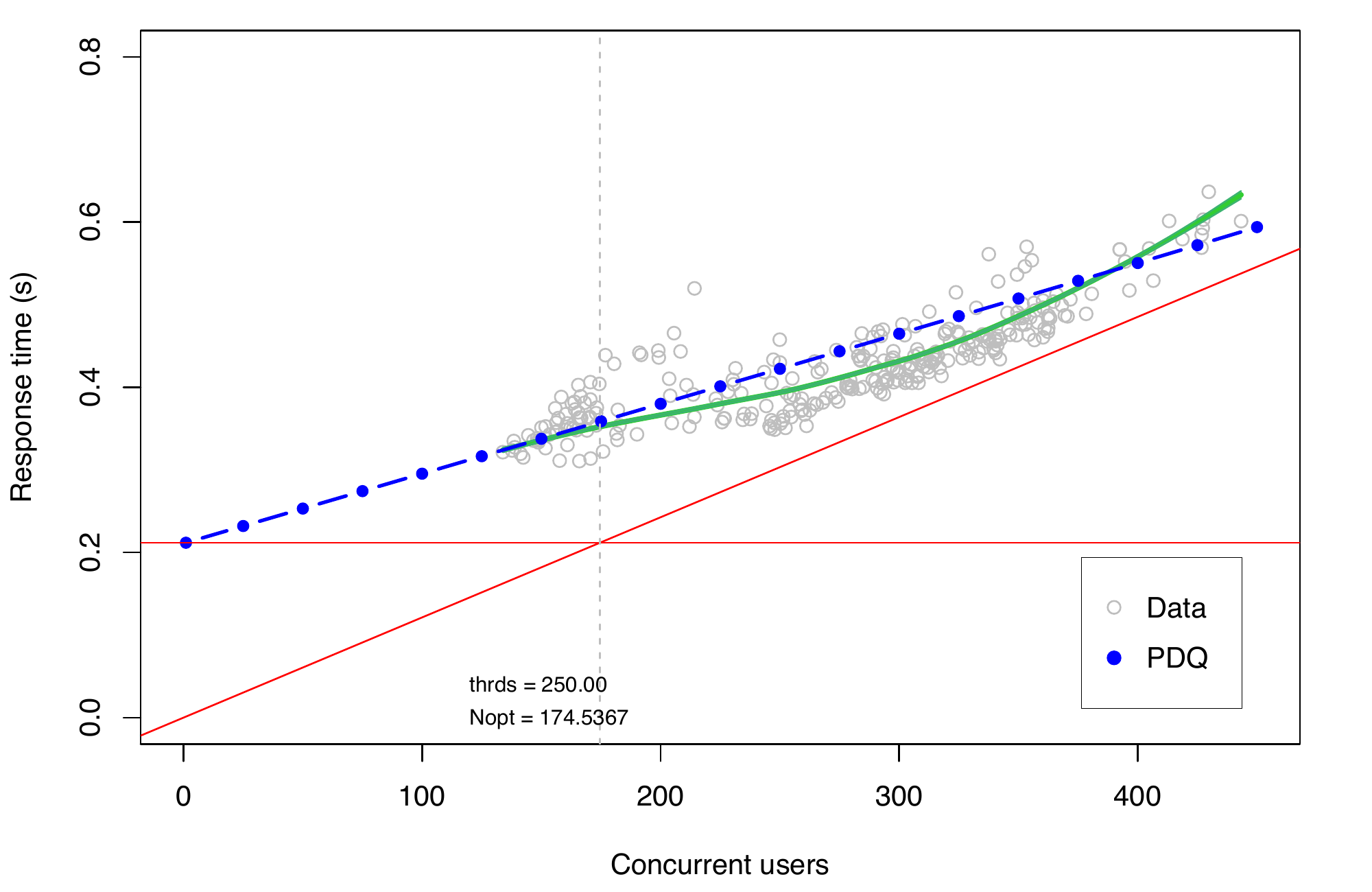}
  \caption{PDQ response time model} \label{fig:pdqR2016jul}
\end{subfigure}
\caption{PDQ performance model (blue dots) of July 2016} \label{fig:pdq2016jul}
\end{figure*}

Figure~\ref{fig:jmxX2016jul} shows the first production data sample: a veritable cloud of data points 
with $N$ in the range of 100 to 450 user requests.
Since they are throughput data, the scatterplot exhibits part of a concave profile, in conformance with the expectations set by Figure~\ref{fig:pdqX}.

The initial PDQ model in Figure~\ref{fig:pdqX2016jul} supports this impression.
And similarly for the corresponding response time data in Figure~\ref{fig:pdqR2016jul}, 
the PDQ model exhibits the expected hockey-stick handle.

Although our general expectations are met, from a visual persepctive, there are some outstanding questions.
The initial PDQ model did not conform to the diagram in Figure~\ref{fig:pdqtc}. 
 
Many additional queues were required to calibrate with the observed minimum response time of $R_{min} = 0.45$ seconds.
The question then became, what did these additional queues represent in the real Tomcat server? 
Two distinct interpretations presented themselves:
\begin{enumerate}
\item Polling external services
\item Hidden parallelism
\end{enumerate}

\noindent
For example, consider the data in the last row of Table~\ref{tab:metrix}.  
The Linux CPU utilization is $U_{dat} = 0.4653$ (or 46.53\%) and the corresponding throughput is 
\mbox{$X_{dat} = 519.75$} requests per second. We can then estimate the average service time using Little's microscopic law in R:
{\small 
\begin{verbatim} 
> X <- 519.748550
> U <- 0.465260
> U / X
[1] 0.0008951636
\end{verbatim}
}
\noindent
That number rounds up to $0.001$, which implies $S_{cpu} = 1$ millisecond.

If a Tomcat thread polled the external services for a response and each polling cycle incurred a service time of 1 millisecond on CPU, a serial chain of some two hundred queues would be needed to match the minimum response time bound 
(horizontal red line) in Figure~\ref{fig:pdqR2016jul}.

The hidden-parallelism interpretation is much more subtle. See the sidebar {\bf \sf Parallel is Just Fast Serial} for an explanation.

The interpretation was not clearly identified until a new set of performance data became available a few months later.
(See Figure~\ref{fig:pdq2016oct})

\section*{\sf PDQ Model Validation} 
Without going into details that would take us too far afield, part of the problem 
described in the previous section 
arose out of the otherwise natural assumption that the service time of a thread was determine by 
its CPU utilization, $U_{dat}$. 

\begin{figure*}[ht]
\centering
\begin{subfigure}{0.33\textwidth}
  \centering
  \includegraphics[scale=0.325]{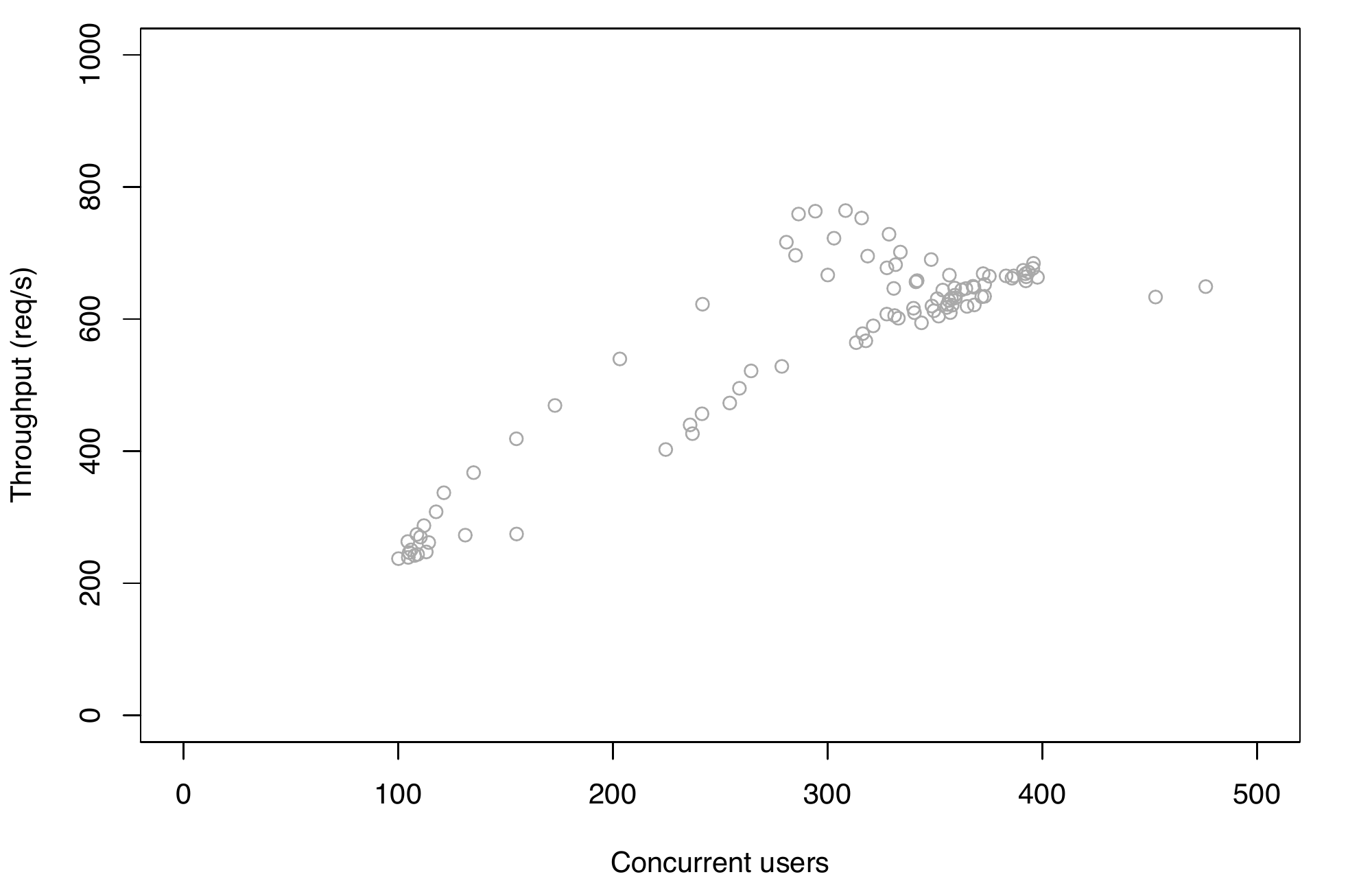}
  \caption{Production throughput data} \label{fig:jmxX2016oct}
\end{subfigure}%
\begin{subfigure}{0.33\textwidth}
  \centering
  \includegraphics[scale=0.325]{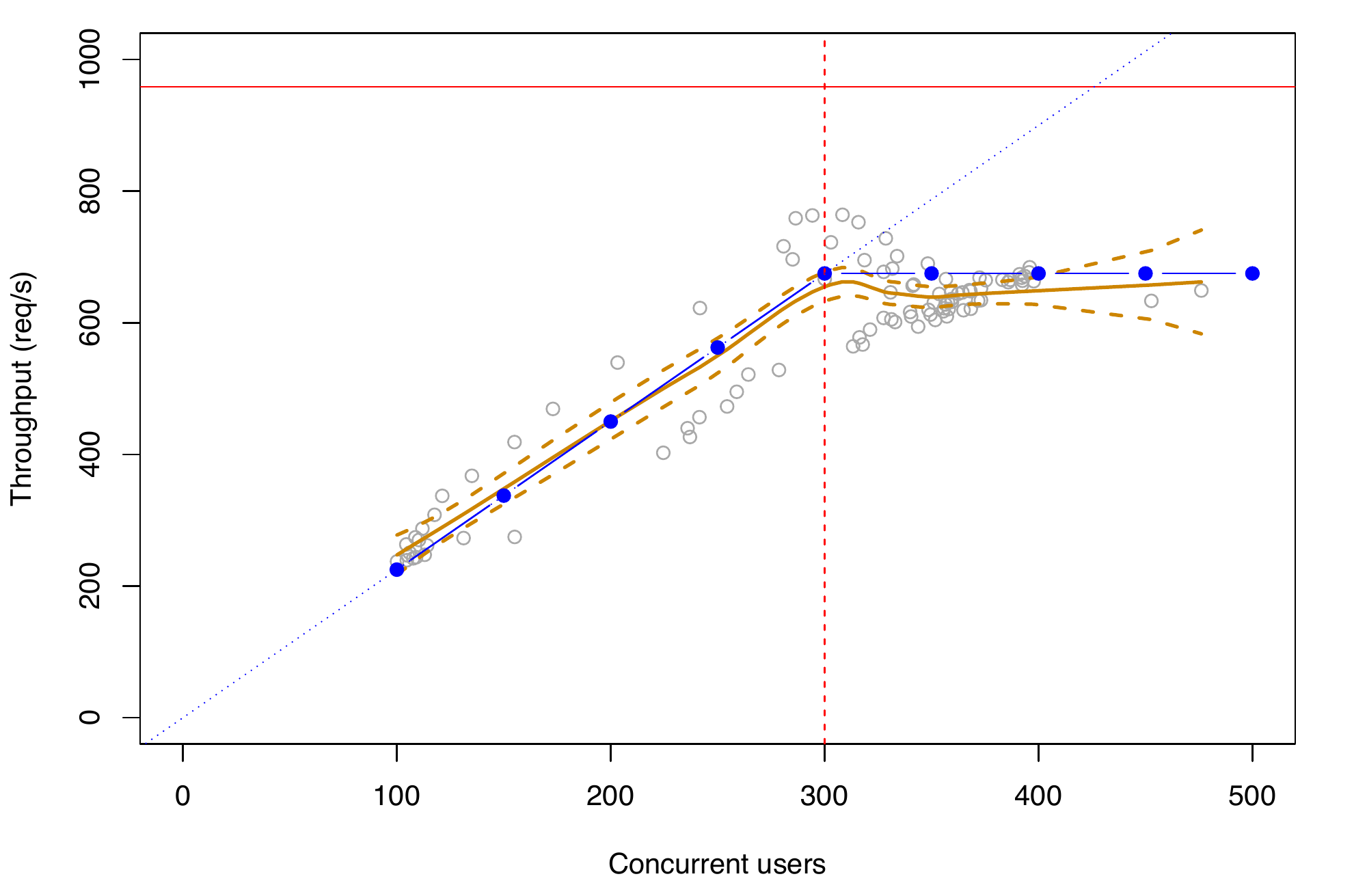}
  \caption{PDQ throughput model} \label{fig:pdqX2016oct}
\end{subfigure}%
\begin{subfigure}{0.33\textwidth}
  \centering
  \includegraphics[scale=0.325]{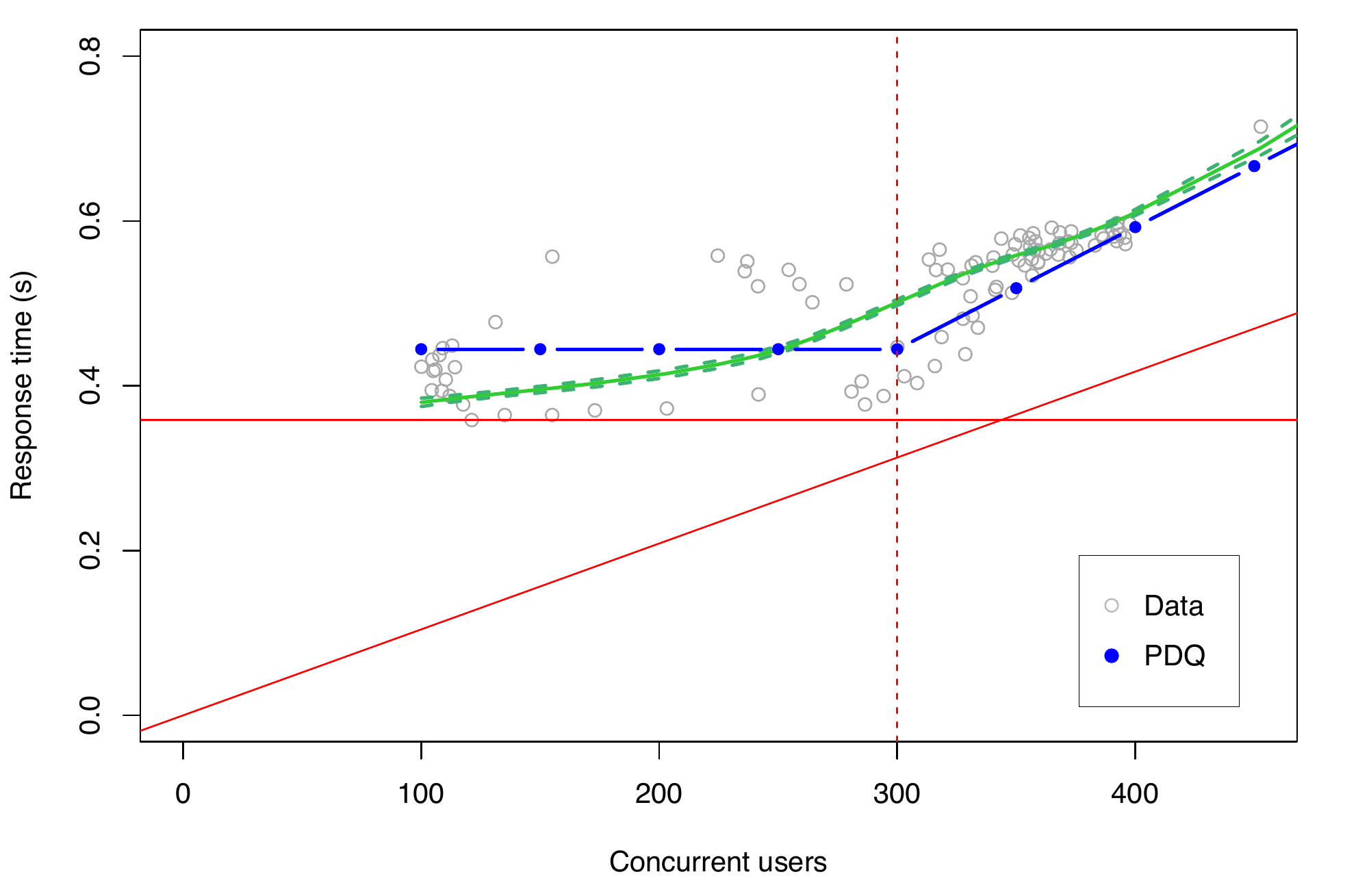}
  \caption{PDQ response time model} \label{fig:pdqR2016oct}
\end{subfigure}
\caption{PDQ performance model (blue dots) of October 2016} \label{fig:pdq2016oct}
\end{figure*}

The new throughput data in Figure~\ref{fig:jmxX2016oct} is sparser than for Figure~\ref{fig:jmxX2016jul} and suggests that the smartphone requests were lighter for that period. This sparseness made it clearer that 
the additional queues were actually a sign of hidden parallelism, viz., linear throughput. 
%(See sidebar {\bf \sf Parallel is Just Fast Serial}) 
And the source of that parallelism is the Tomcat threads themselves. That is how it works! 

Up to a certain threshold (to be discussed shortly), every arriving smartphone request is immediately assigned to a Tomcat thread and they all run simultaneously---in parallel. 

This dynamic produces a subtle change in the steady-state throughput profile. 
Parallel throughput is a special case of Figure~\ref{fig:pdqX} where, although the functional shape of $X$ is still concave (as described earlier), it not curved. Rather, it is linear rising all the way up to saturation, whereupon it immediately flattens out to a plateau. This difference can be seen very clearly in Figure~\ref{fig:pdqX2016oct}. 
The knee point in these data occurs at about $N_{knee}=300$ threads. 
The plateau beyond that knee corresponds to constant throughput.

More importantly, the blue dots in Figure~\ref{fig:pdqX2016oct} are produced by the PDQ model in 
Figure~\ref{fig:pdqtc} and they now precisely reflect the average behavior of the throughput data. 
Similarly for the PDQ model (blue dots) of the response time data in Figure~\ref{fig:pdqR2016oct} where we now see, not only the hockey-stick handle but, the foot of the hockey-stick below the knee point.

\begin{figure*}[ht]
\centering
\includegraphics[scale=0.6]{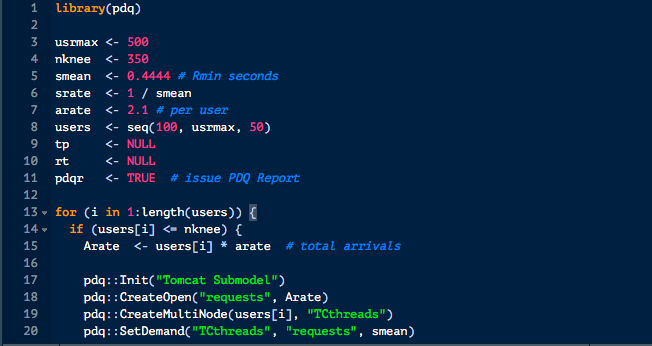}
\caption{\sf Portion of PDQ Tomcat model written in the R language}  \label{fig:pdqcode}
\end{figure*}

\begin{figure*}[ht]
\centering
\begin{subfigure}{0.5\textwidth}
  \centering
  \includegraphics[scale=0.6]{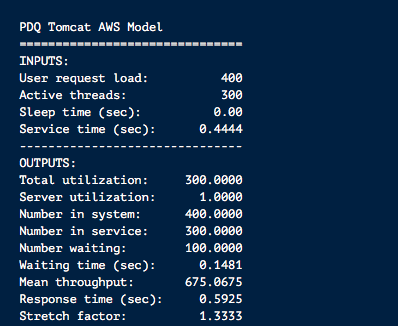}
  \caption{PDQ parameters for 2016 data} \label{fig:pdqout16}
\end{subfigure}%
\begin{subfigure}{0.5\textwidth}
  \centering
  \includegraphics[scale=0.6]{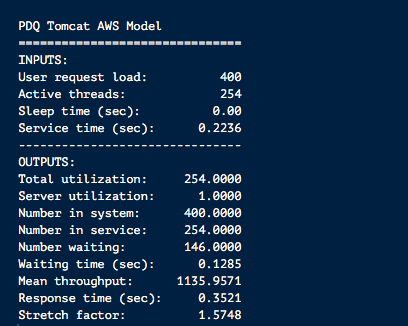}
  \caption{PDQ parameters for 2018 data} \label{fig:pdqout18}
\end{subfigure}
\caption{Example PDQ modeling outputs}  \label{fig:pdqouts}
\end{figure*}

This led to the following adjustments in the original PDQ model code in Figure~\ref{fig:pdqcode}: 
\begin{itemize}
\item The correct service time, $S_{TC}$, is the total time for which a Tomcat thread is servicing a request.
\item $S_{TC}$ is completely dominated by time spent in the external services, i.e., \mbox{$S_{TC} = R_{min}$}. 
\item From the new data we see that \mbox{$R_{min} = 444.4$} milliseconds in Figure~\ref{fig:pdqout16}. 
\item Parallel execution of Tomcat threads is correctly represented by the PDQ model in Figure~\ref{fig:pdqtc}.
\end{itemize}

Naturally, the previous performance data of July 2016 can be reevaluated using this revised PDQ model, but in the interest of space, we omit those details. 

There is a certain irony in that the development of the first PDQ model took a ``wrong turn'' due to having   
too much data that ``clouded'' the $X(N)$ and $R(N)$ profiles.

The reader should note that this kind of confusion is the rule rather than the exception in this kind of performance analysis. There is no right or wrong, just progress.

\section*{\sf PDQ Model Predictions}
With a validated PDQ model in place, we can use it to explore the performance of various AWS cloud configuration options.  %with the aim of reducing our Amazon service fees. 
In general, this can be an painstaking process that is application-dependent, so we only present our key findings.

As mentioned at the outset, A/S groups were part of the original architecture in Figure~\ref{fig:apparch}. 
In particular, the A/S policy is triggered when a Linux-Tomcat instance causes the CPUs to exceed 75\% busy, i.e., 
$U_{cpu} \geq 0.75$.
A/S then spins up additional EC2 instances to shed the increasing smartphone traffic via the ELB.

In order to maintain the $U_{cpu}$ threshold, in any instance, 
no additional Tomcat threads are invoked above that threshold, which corresponds to 
\mbox{$N_{knee} = 254$} threads in Figure~\ref{fig:pdq2016jul} and
\mbox{$N_{knee} = 300$} threads in Figure~\ref{fig:pdq2016oct}.
As described earlier, this causes the instance throughput to plateau beyond the vertical knee-point line in 
Figures~\ref{fig:pdqX2016oct} and~~\ref{fig:pdqR2016oct}.

The reader should note that this saturation effect is not due to native Linux. 
The Linux time-share scheduler is not designed to throttle processor utilization at some arbitrary value below 100\% 
(i.e., processor saturation).

On the contrary, it is only A/S that can induce pseudo-saturation knee point.

As discussed earlier, this kind of resource constraint is a feature of a fair-share scheduler,  
and that means it can only be achieved via the hypervisor (Figure~\ref{fig:xenarch}) or related code in AWS or 
something like c-groups~\cite{cgroups} in Linux.

\begin{figure*}[ht]
\centering
\includegraphics[scale=0.65]{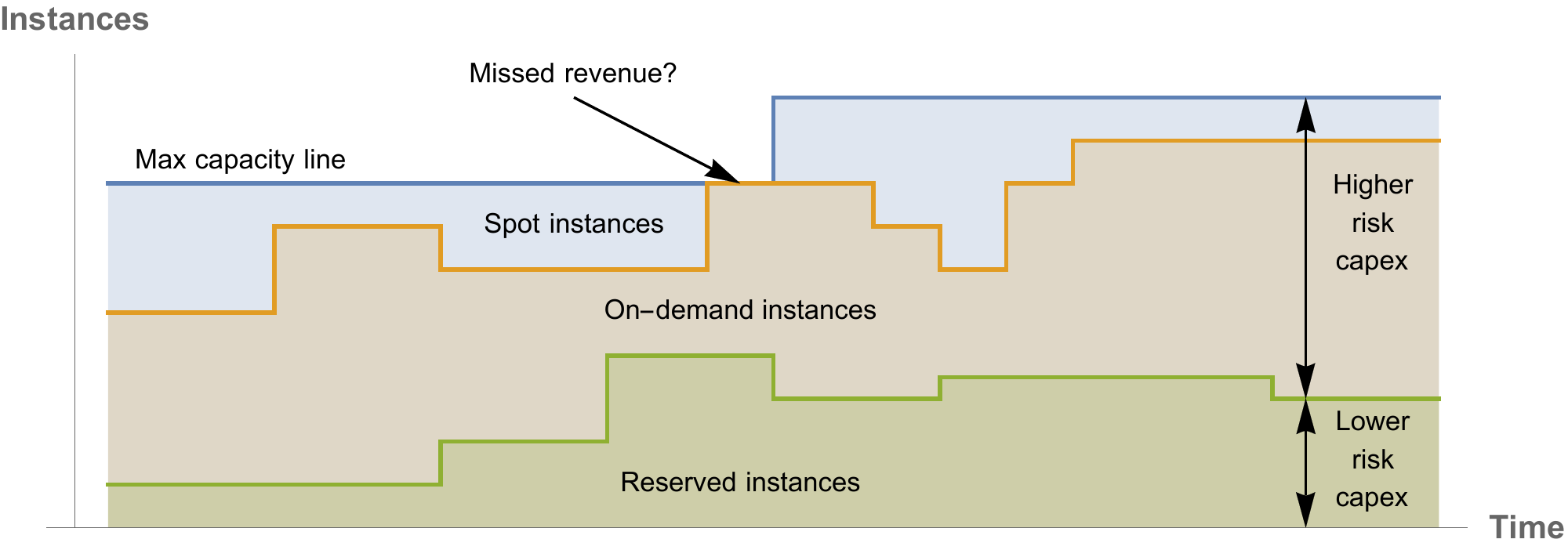}
\caption{AWS infrastructure capacity lines (Source~\cite{cloud2})}  \label{fig:awscap} 
\end{figure*}

There are a variety of hidden costs associated with A/S.  
The A/S policy threshold based on processor utilization is no necessarily the most efficient capacity criterion. 
There is also typically delay of about 10 minutes for additional EC2 instances to spin up. 

Based on our validated PDQ model, we considered employing pre-emptive EC2 scheduling.
This Scheduled Scaling (S/S) policy is indicated by the {\em clock} in Figure~\ref{fig:awssched}.
S/S is cheaper than A/S with a typical savings of about 10\% in AWS fees. 

When using S/S, the number of concurrent Tomcat service threads, $N$, should be used 
to size the number of EC2 instances required for incoming smartphone traffic.
The use of S/S also removes observed spikes in both arriving traffic into an EC2 instance and the 
associated EC2 spin-up delay.

An even more cost-effective solution is presented by Amazon Spot Instances (S/I), shown schematically in Figure~\ref{fig:awsspot}. S/Is are available at 90\% discount on standard on-demand pricing. 
The S/I offering is motivated by Amazon trying to recover the cost of its own infrastructure capacity~\cite{cloud2}. 

As Figure~\ref{fig:awscap} shows, Reserved Instances represent the easiest capacity for Amazon to plan.
The needed capacity is well-known.
On-demand capacity represents higher risk for Amazon because not all of it may be consumed in the way anticipated.
S/I capacity is what is left over between the expected demand and maximum possible physical capacity and therefore 
represents the highest risk for Amazon. Demand for that unused headroom is created by Amazon offering 
it at an extremely large discount to customers.

On the other hand, it can be challenging for the system administrator to diversify instance types and sizes within the same group. For example, the default instance type might be {\tt m4.10xlarge}, 
whereas the S/I market may only be offering the smaller {\tt m4.2xlarge} instance type. 
This situation can force manual reconfiguration of the application. 

\begin{figure*}[ht]
\centering
\begin{subfigure}{0.5\textwidth}
  \centering
  \includegraphics[scale=0.4]{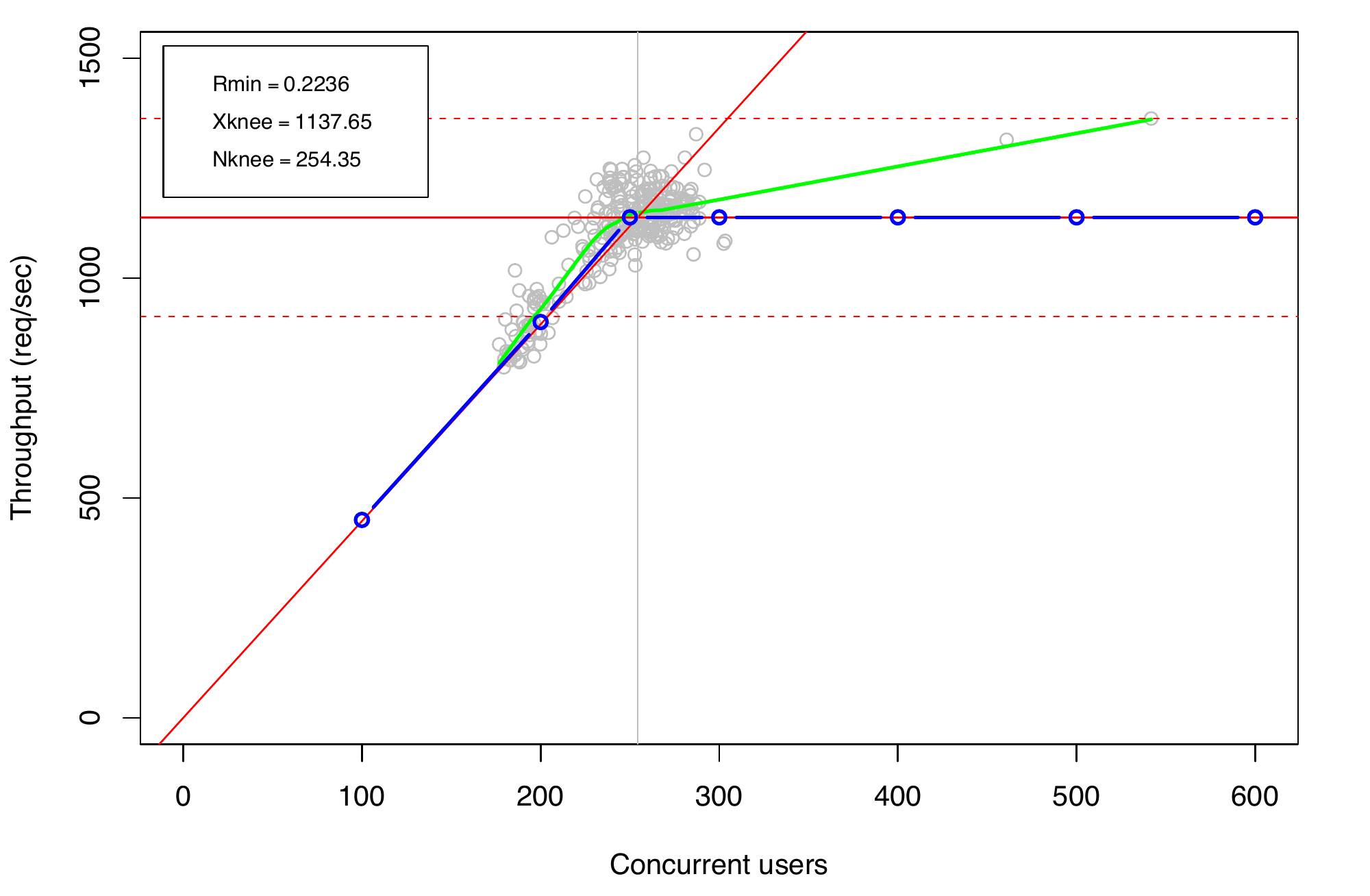}
  \caption{Production throughput} \label{fig:pdqX2018mar}
\end{subfigure}%
\begin{subfigure}{0.5\textwidth}
  \centering
  \includegraphics[scale=0.4]{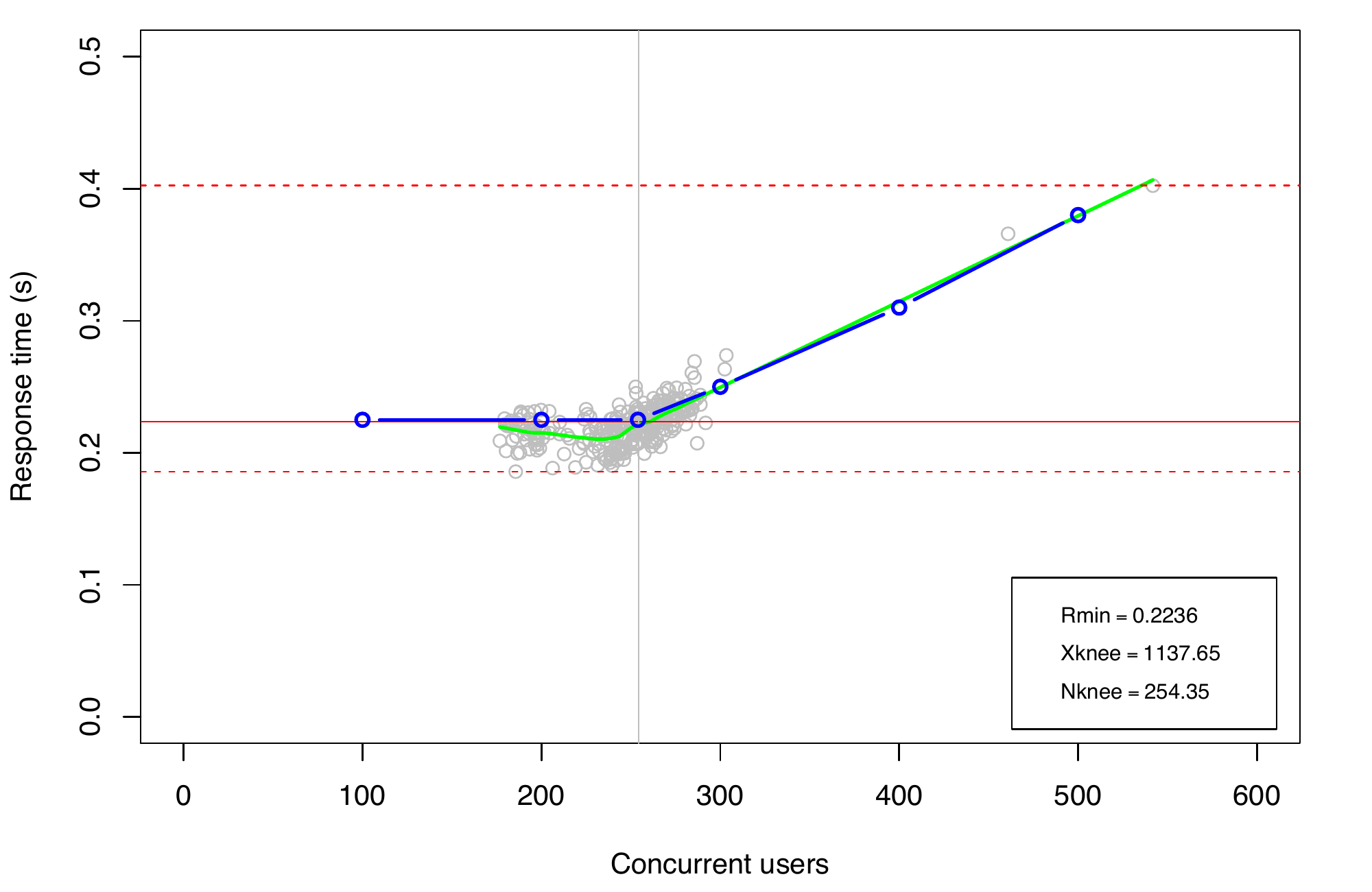}
  \caption{Production latency} \label{fig:pdqR2018mar}
\end{subfigure}
\caption{PDQ performance model (blue dots) for March 2018 data} \label{fig:pdq2018mar}
\end{figure*}

Finally, the overall payoff for doing this kind of performance modeling can be seen in Figure~\ref{fig:pdq2018mar}.
The range in concurrent requests has been narrowed to between 200 and 300. There are only a few excursions beyond the  
pseudo-saturation knee point, which has now moved down slightly to $N_{knee} = 254$ threads.  

Additionally, the minimum response time has been reduced to \mbox{$R_{min} = 0.2236$} seconds and the concomitant 
system throughput has increased to \mbox{$X_{max} = 1135.96$} requests per second. 
These changes are reflected in the PDQ model outputs shown in Figure~\ref{fig:pdqout18}.

\section*{\sf Wait Where?}
Constantly checking  consistency is an essential part of all performance analysis. 
One question that still remains outstanding is, 
where do user requests wait when their number is above the A/S knee point? In other words, where are the ``blocks'' in Figure~\ref{fig:pdqtc} located in the actual Tomcat system? 

We know requests must wait because PDQ tells us so. 
When $N$ exceeds $N_{knee}$, no more Tomcat threads can be allocated, so  
a new request must wait for an already busy service thread to become available. But wait where?

Further support for this question comes from the production data in 
Figures~\ref{fig:pdqR2016oct} and~\ref{fig:pdqR2018mar}. The hockey-stick handle is clearly visible and queueing theory tells us the handle corresponds to a growing waiting line in Figure~\ref{fig:pdqtc}.

At any instant, a JVM thread can be in one of the following states~\cite{oracle}:
\begin{description}
\item[\sc new: ] A thread that has not yet started is in this state.
\item[\sc runnable: ] A thread executing in the Java virtual machine is in this state.
\item[\sc blocked: ]
A thread that is blocked waiting for a monitor lock is in this state.
\item[\sc waiting: ]
A thread that is waiting indefinitely for another thread to perform a particular action is in this state.
\item[\sc timed\_waiting: ]
A thread that is waiting for another thread to perform an action for up to a specified waiting time is in this state.
\item[\sc terminated: ]
A thread that has exited is in this state.
\end{description}

\noindent
The following instantaneous samples
\begin{quote}
{\footnotesize
\begin{verbatim}
TIMED_WAITING    RUNNING
138              301
111              306
173              519
108              286
 65              152
 68              185
 72              119
\end{verbatim}%
}
\end{quote}

\noindent
indicate that threads are always waiting despite PDQ telling us there should be no waiting below $N_{knee}$.

Further investigation revealed a clash between nomenclatures. JVM waiting states have a different meaning from the queue-theoretic waiting states. 
In the context of JVM threads, ``waiting'' refers to a thread waiting for {\em work}~\cite{stackjvm}, 
whereas in our PDQ models, requests wait for service and a thread server ``waiting'' for work is {\em idle}. 

That would explain why we see non-zero waiting values below $N_{knee}$.
Once a thread has finished executing on behalf of a user request, there can be some small delay to reinitialize that thread before it can accept a new request. 

With all this in mind, our best guess is that requests that are waiting in the PDQ model are actually waiting for service in the CPU run-queue of the Linux kernel.

%%%%%%%%%%%%   Parallel = Fast Serial sidebar %%%%%%%%%
\begin{figure*}[ht]
\centering
\fbox{
\begin{minipage}{0.75\textwidth}
{\bf \sf \large Parallel is Just Fast Serial}\\
\footnotesize \sf 
From the standpoint of queueing theory, parallel processing can be regarded as a form of fast serial processing. 
The left side of the diagram shows a pair of parallel queues, where requests arriving from outside at rate $\lambda$ are split equally to arrive with reduced rate $\lambda/2$ into either one of the two queues. 
Assume $\lambda=0.5$ requests/second and $S=1$ second.
\begin{center}
\includegraphics[scale=0.35]{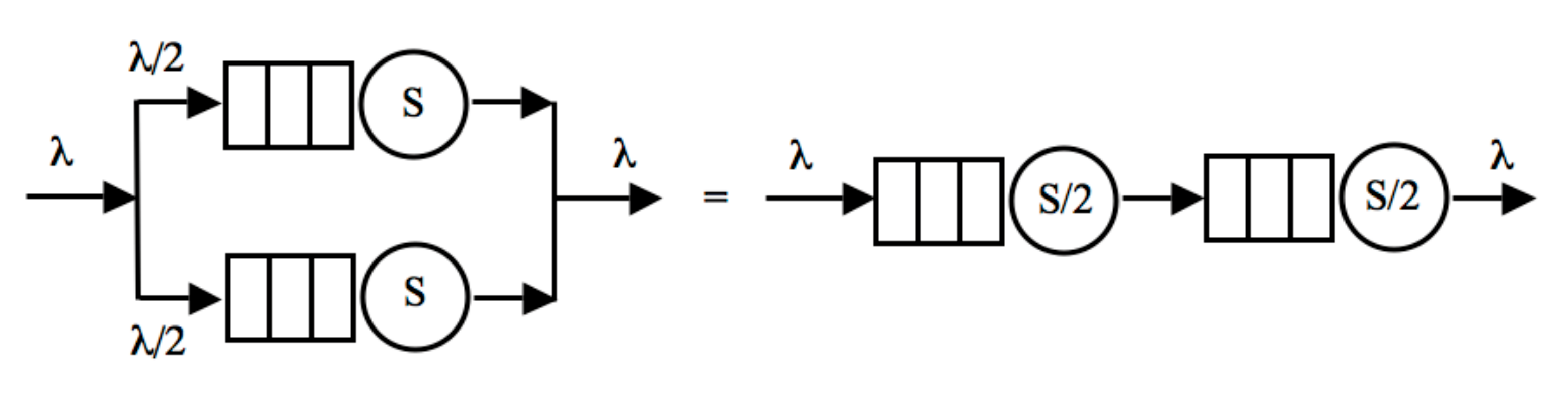}
\end{center}
When a request joins the tail of one of the parallel waiting lines, its expected time to get through that queue (waiting + service) is given by equation (1) in {\em Berechenbare Performance}~\cite{berechenbare}, namely:
\begin{equation}
T_{para} \; = \quad \dfrac{S}{1 - (\lambda/2) S} \quad = \; 1.33~\text{seconds} \label{eqn:para}
\end{equation}
%The right side of the diagram shows two queues in tandem, each having a faster service time, $S/2$.
The right side of the diagram shows two queues in tandem, each twice as fast ($S/2$) as a parallel queue. 
Since the arrival flow is not split,  
the expected time to get through both queues is the sum of the times spent in each queue:
\begin{equation}
T_{serial} \; = \quad \dfrac{S/2}{1 - \lambda (S/2)} + \dfrac{S/2}{1 - \lambda (S/2)} \quad = \quad \dfrac{S}{1 - \lambda (S/2)} \quad = \;  1.33~\text{seconds}  \label{eqn:serial}
\end{equation}
$T_{serial}$ in equation~\eqref{eqn:serial} is identical to $T_{para}$ in equation~\eqref{eqn:para}.
Conversely, multi-stage serial processing can be transformed into an equivalent form of 
parallel processing~\cite{pdqbook,pdqtool}. This insight helped identify the ``hidden parallelism'' in the July and October 2016 performance data that led to the  
correction of the initial PDQ Tomcat model.
\end{minipage}
}
\end{figure*}
%%%%%%%%%%%% %%%%%%%%%%%% %%%%%%%%%%%% 

\section*{\sf Conclusion}
Cloud services are more about economic benefit for the service provider than they are about technological innovation 
for the user. It is not merely plug-and-play but pay-and-pay!
It therefore becomes incumbent on application architects and system administrators to minimize the cost of cloud services for their organization.

Meaningful cost-benefit decisions can only be made with the aid of ongoing performance analysis and capacity planning. 

In this article, we have shown how PDQ models can provide insight into cloud-based sizing and performance. 
The queueing model framework helps expose where hidden performance costs actually reside and what to do about them. 

The introduction of AWS Lambda~\cite{cloud2} is likely to introduce new modifications into our future PDQ models.  
We also did not attempt to take into account so-called ``grey failure'' performance anomalies~\cite{graycloud}, 
which could be part of future work.

\hrulefill
%%%%%%% REFERENCES %%%%%%%%%%
\renewcommand{\refname}

%%%%%%% Authors %%%%%%%%%%
\hrulefill
\section*{\sf Authors}
Neil Gunther, M.Sc., Ph.D., is an internationally recognized computer scientist who
founded Performance Dynamics in 1994. Prior to that, he 
held positions at San Jose State University, NASA-JPL, Xerox PARC and Pyramid-Siemens Corp. 
Dr. Gunther is a senior member of ACM and IEEE, and received the A.A. Michelson Award in 2008.

Mohit Chawla is an independent systems engineer based in Hamburg, Germany.
%%%%%%% Authors %%%%%%%%%%

\end{multicols}
\end{document}